\documentclass[%
 reprint,
 amsmath,amssymb,
 aps,
]{revtex4-2}

\usepackage{graphicx}
\usepackage{dcolumn}
\usepackage{bm}
\usepackage[colorlinks, allcolors=blue]{hyperref}
\usepackage{soul}
\usepackage{color}

\begin{document}

\preprint{APS/123-QED}

\title{Transmission efficiency limit for nonlocal metalenses}

\author{Shiyu Li}

\affiliation{Ming Hsieh Department of Electrical and Computer Engineering, University of Southern California, Los Angeles, California 90089, USA}

\author{Chia Wei Hsu}
\email{cwhsu@usc.edu}
\affiliation{Ming Hsieh Department of Electrical and Computer Engineering, University of Southern California, Los Angeles, California 90089, USA}

\begin{abstract}
The rapidly advancing capabilities in nanophotonic design are enabling complex functionalities limited mainly by physical bounds.
The efficiency of transmission is a major consideration, but its ultimate limit remains unknown for most systems.
Here, 
we introduce a matrix formalism that puts a fundamental bound on the channel-averaged transmission efficiency of any passive multi-channel optical system based only on energy conservation and the desired functionality,
independent of the interior structure and material composition.
Applying this formalism to diffraction-limited nonlocal metalenses with a wide field of view,
we show that the transmission efficiency must decrease with the numerical aperture for the commonly adopted designs with equal entrance and output aperture diameters.
We also show that reducing the size of the entrance aperture can raise the efficiency bound.
This work reveals a fundamental limit on the transmission efficiency as well as providing guidance for the design of high-efficiency multi-channel optical systems. 
\end{abstract}

\maketitle


\section{Introduction}
\label{sec:intro}

Over the past decade, nanophotonic design and fabrication became more and more advanced.
Oftentimes, what limits the device performance is no longer fabrication constraints or the cleverness of the design, but fundamental physical bounds.
Furthermore, the design process typically requires time consuming development, simulation, and optimization.
It is invaluable to know beforehand what the fundamental bounds are and how they are related to the design choices~\cite{chao2022physical}.
Such a knowledge can significantly reduce the time spent in blind explorations and also point to better design choices that are not otherwise obvious.
Of particular interest are multi-channel optical systems.
For example, with a metalens~\cite{lalanne2017metalenses,khorasaninejad2017metalenses,liang2019high,pan2022dielectric,arbabi2023advances}, one would like the incident wave from each angle to be focused to the corresponding focal spot with unity efficiency.
But is a uniformly perfect efficiency compatible with the desired angle-dependent response?
What is the highest efficiency allowed by fundamental laws?


Here, we introduce a matrix-based formalism that sets a fundamental bound on the efficiency of any linear multi-channel system given its functionality, and apply it to metalenses.

Metalenses are compact lenses made with metasurfaces, 
which show great potential for thinner and lighter imaging systems with performances comparable to or exceeding conventional lenses~\cite{lalanne2017metalenses,khorasaninejad2017metalenses,liang2019high,pan2022dielectric,arbabi2023advances}.
Metalenses designed from a library of unit cells have limited focusing efficiency, which can be overcome by more flexible designs~\cite{chung2020high,schab2022upper}.
Inverse design~\cite{sell2017periodic,pestourie2018inverse,liang2018ultrahigh,mansouree2018large,phan2019high,lin2019achieving,zhuang2019high,chung2020high,schab2022upper}, grating averaging technique~\cite{arbabi2020increasing}, and stitching separately designed sections together~\cite{byrnes2016designing,paniagua2018metalens,kang2020efficient} are effective approaches.
However, achieving high focusing efficiency at large numerical aperture (NA) remains difficult,
as all such ``local'' metasurfaces have limited deflection efficiency at large angles~\cite{estakhri2016wave,asadchy2016perfect}.
Since local metasurfaces have a spatial impulse response close to a delta function, they provide the same response for different incident angles, so they are also limited in their angular field of view (FOV)~\cite{li2022thickness}.

Nonlocal metalenses with tailored interactions between adjacent building blocks ({\it i.e.}, the spatial impulse response is extended beyond a delta function) can overcome the limited angular diversity of local metalenses to enable diffraction-limited focusing over a large FOV~\cite{overvigdiffractive,li2022thickness,shastri2022nonlocal}.
Nonlocal metalenses based on doublets~\cite{arbabi2016miniature,groever2017meta} or aperture stops~\cite{shalaginov2020single,fan2020ultrawide} can support focusing efficiencies higher than 50\% over a wide FOV, but with NA lower than 0.5.
Multi-layer structures obtained from inverse design have achieved diffraction-limited focusing with NA = 0.7 over FOV = 80$^{\circ}$, but the averaged focusing efficiency is only about 25\%~\cite{lin2021computational}.
However, there was no guidance on the efficiency bound of these nonlocal metalenses.

\begin{figure*}[t]
\centering
\includegraphics[width=0.95\textwidth]{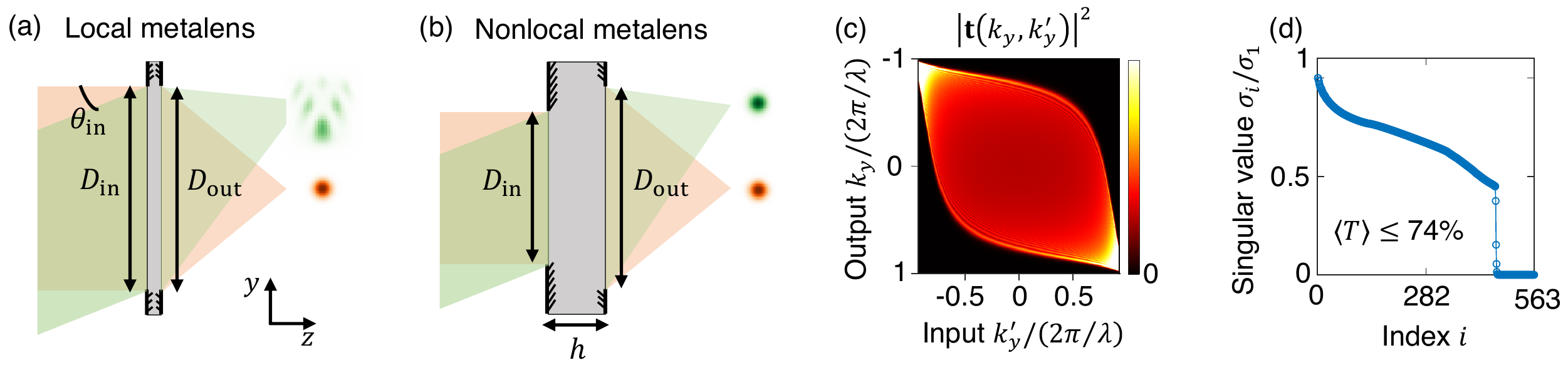}
\caption{Nonlocal metalenses and their transmission matrix. (a) A local metalens with subwavelength thickness, for which the input and output apertures have the same size, {\it i.e.}, $D_{\rm in}=D_{\rm out}$. Diffraction-limited focusing can only be achieved within a narrow range of incident angles.
(b) A nonlocal metalens, whose larger thickness allows nonlocal coupling, under which diffraction-limited focusing may be achieved over a wide angular range. Here, $D_{\rm in}$ and $D_{\rm out}$ can be different.
(c,d) The squared amplitude of the transmission matrix $|\mathbf{t}(k_y,k_y')|^2$ (c) and its normalized singular values $\sigma_i /\sigma_1$ (d) for an ideal nonlocal metalens with diameter $D_{\rm in}=D_{\rm out}=300\lambda$, NA = 0.8, FOV = 140$^{\circ}$. Here, the averaged transmission efficiency is bounded by \eqref{eq:TE} as $\langle T \rangle \leq 74\%$.}
\label{fig:Fig1}
\end{figure*}

From the desired response of a multi-channel optical system, we can write down its transmission matrix that relates the input to the output.
Here, we rigorously bound the channel-averaged transmission efficiency using the singular values of the transmission matrix and the fact that the transmitted energy must not exceed the input energy.
For commonly adopted designs with equal entrance and output apertures, we find the transmission efficiency bound of a nonlocal metalens to drop with the NA. 
We also find that reducing the entrance aperture size can raise the efficiency bound to close to unity.
This approach is general and can guide the design of not only metalenses but also other multi-channel optical systems.

\section{Results}
\label{sec:results}

\subsection{Nonlocality}\label{subsec:nonlocality}

As schematically illustrated in Fig.~\ref{fig:Fig1}(a),
subwavelength-thick local metalenses (such as metalenses with the hyperbolic phase profile~\cite{aieta2012aberration}) perform well only over a limited input angular range.
Nonlocal metalenses can achieve diffraction-limited focusing over a much wider FOV but need a minimal thickness to provide the nonlocality~\cite{li2022thickness, 2023_Miller_Science}.
Since a large FOV requires an angle-dependent response ({\it i.e.}, angular dispersion), a spatially localized incident wave must spread as it propagates through the metalens under space-angle Fourier transform.
Thus more angular diversity necessitates more nonlocality.
As shown in Supplementary Fig.~S1, nonlocal effects become important when the FOV is larger than a threshold.
For FOV above such a threshold, a sufficient thickness is needed for light to spread and create nonlocality,
so the input diameter $D_{\rm in}$ and output diameter $D_{\rm out}$ can be different [Fig.~\ref{fig:Fig1}(b)].
The different aperture sizes provide an additional degree of freedom compared to local metalenses with subwavelength thicknesses (for which $D_{\rm in}=D_{\rm out}$ intrinsically).

\subsection{Transmission matrix and target response}

While the transmission-matrix formalism introduced in this work is general, for concreteness we will consider metalenses.
For ideal (aberration-free) lenses under plane-wave incidence,
we can obtain the transmitted phase profile on the output surface of the metalens by matching the optical path lengths between the marginal rays and the chief ray at the focus ${\bf r}(\theta_{\rm in}) = (y=f\tan{\theta_{\rm in},z=f+h})$. In 2D, we get~\cite{kalvach2016aberration,shalaginov2020single,li2022thickness}
\begin{equation}
    \phi_{\rm out}^{\rm ideal}(y,z=h,\theta_{\rm in}) =
     \psi(\theta_{\rm in})-\frac{2\pi}{\lambda}\sqrt{f^2+\left (y-f\tan{\theta_{\rm in}} \right )^2},
     \label{eq:phase_2D}
\end{equation}
where $\theta_{\rm in}$ is the incident angle; $f$ and $h$ are the focal length and lens thickness respectively; $\psi(\theta_{\rm in})$ is an angle-dependent but spatially invariant global phase with no effect on the focusing quality.

Consider monochromatic, transverse magnetic waves of 2D systems at wavelength $\lambda$, where ${\bf E}=E_x(y,z)\hat{x}$.
The incoming wavefront $E_x(y',z=0)$ can be projected onto $N_{\rm in}$ propagating plane waves with input transverse momenta $k_y'=\{k_y^a\}=(2\pi/\lambda)\sin{\theta_{\rm in}}$ truncated to the size of the input aperture {\it i.e.}, $|y'|<D_{\rm in}/2$, yielding complex-valued amplitude $\alpha_a$ of the $a$-th plane-wave input, where $z=0$ is the front surface of the system,
$|k_y^a|<(2\pi/\lambda)\sin{(\rm FOV/2)}$ is restricted to the FOV of interest with $k_y^a$ discretized with $2\pi/D_{\rm in}$ spacing following the Nyquist-Shannon theorem~\cite{landau1967sampling}.
Similarly, the transmitted wavefront $E_x(y,z=h)$ can be written as a superposition of $N_{\rm out}$ plane waves with $k_y=\{k_y^b\}=(2\pi/\lambda)\sin{\theta_{\rm out}}$ truncated to $|y|<D_{\rm out}/2$, and the complex-valued amplitude of the $b$-th plane-wave output is $\beta_b$, where $\theta_{\rm out}$ is the output angle and $|k_y^b|<2\pi/\lambda$ is sampled with momentum spacing $2\pi/D_{\rm out}$.
For any linear optical system, these amplitudes are related through the transmission matrix $t_{ba}$,
\begin{equation}
   \beta_b = \sum_{a=1}^{N_{\rm in}}t_{ba}\alpha_a.
   \label{eq:transmisson_matrix}
\end{equation}
We can parameterize the incoming and transmitted wavefronts as column vectors ${\bm \alpha}=[\alpha_1;\cdots;\alpha_{N_{\rm in}}]$ and ${\bm \beta}=[\beta_1;\cdots;\beta_{N_{\rm out}}]$. Then, ${\bm \beta}=\mathbf{t}{\bm \alpha}$,
where $\mathbf{t}=\mathbf{t}(k_y,k_y')=t(k_y^b,k_y^a)=t_{ba}$ is the transmission matrix in the angular basis.

Fourier transforming the output field of the ideal metalens of Eq.~(\ref{eq:phase_2D}) from $y$ to $k_y$ yields the transmission matrix $\mathbf{t}$.
Such transmission matrix applies to any lens with diffraction-limited focusing over this FOV.
The squared amplitude of the transmission matrix $|\mathbf{t}(k_y,k_y')|^2$ for a metalens with $D_{\rm in}=D_{\rm out}=300\lambda$, ${\rm NA} = \sin(\arctan(D_{\rm out}/(2f))) = 0.8$ and FOV = 140$^{\circ}$ is shown in Fig.~\ref{fig:Fig1}(c).
More details on the transmission matrix can be found in Supplementary Sec.~2. 
Notably, Eq.~(\ref{eq:phase_2D}) does not specify the overall amplitude prefactor of the transmission matrix, so we cannot directly assess the transmission efficiency based on this target response.

\begin{figure*}[tb]
\centering
\includegraphics[width=0.8\textwidth]{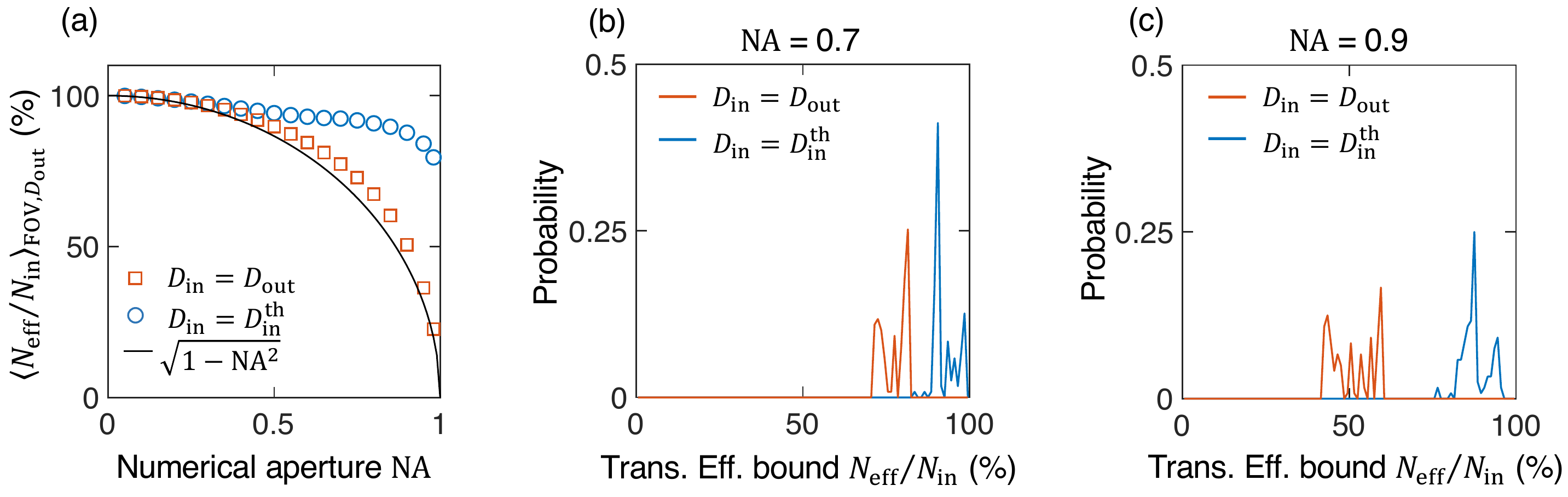}
\caption{Transmission efficiency bound $N_{\rm eff}/N_{\rm in}$ of ideal nonlocal metalenses when $D_{\rm in}=D_{\rm out}$ and when using $D_{\rm in}=D_{\rm in}^{\rm th}$ from \eqref{eq:Din_th}.
(a) $N_{\rm eff}/N_{\rm in}$ averaged over FOV and output diameter $D_{\rm out}$ as a function of NA. The black line is $\sqrt{\rm 1-NA^2}$.
(b,c) The distribution of $N_{\rm eff}/N_{\rm in}$ among different FOV and $D_{\rm out}$ when (b) NA = 0.7 and (c) NA = 0.9.}
\label{fig:Fig2}
\end{figure*}

\subsection{Transmission efficiency bound}

When the transmission matrix includes the overall amplitude prefactor and is flux-normalized, the total transmission efficiency (summed over all transmitted flux) averaged over inputs within the prescribed FOV is $\langle T\rangle = {\sum_{a}T_{a}}/{N_{\rm in}}$,
where $T_a=\sum_{b}|t_{ba}|^2$ is the total transmission efficiency for incident angle $\theta_{\rm in}^a$ and
$N_{\rm in}$ is the number of input channels
({\it i.e.}, the number of columns in the transmission matrix).
To obtain an upper bound on $\langle T\rangle$ without knowledge of the overall amplitude prefactor of the transmission matrix, we consider the singular values $\{\sigma_i\}$ of the transmission matrix.
With the singular value decomposition, the transmission matrix is factorized into $\mathbf{t}(k_y,k_y') = \mathbf{U}\bm{\Sigma}\mathbf{V^{\dagger}}=\sum_i \sigma_i (u_i v_i^{\dagger})$, where each right-singular column vector $v_i$ is a normalized incident wavefront being a linear superposition of the input channels.
The corresponding transmitted wavefront is $\sigma_i u_i$, with a transmission efficiency of $\sigma_i^2$ and with the normalized transmitted wavefront being column vector $u_i$.
Energy conservation imposes that the transmitted energy cannot exceed the input energy, so $0 \le \sigma_i^2 \le 1$ for all $i$.
Since 
tr($\bf t^{\dagger}t$) =$\Sigma_a({\bf t^{\dagger}t})_{aa}=\sum_{b,a}|t_{ba}|^2$ equals the sum of the eigenvalues $\lambda_i=\sigma_i^2$ of matrix $\bf t^{\dagger}t$, 
we have $ \langle T\rangle=\sum_{b,a}|t_{ba}|^2/{N_{\rm in}} = {\sum_{i} \sigma_i^2}/{N_{\rm in}}$.
In addition, $\sum_{i} \sigma_i^4 \le \sum_{i} \sigma_i^2$ follows from energy conservation. 
Therefore, we have a rigorous inequality
\begin{equation}
    \langle T\rangle \equiv \frac{\sum_{b,a}|t_{ba}|^2}{N_{\rm in}}
    = \frac{N_{\rm eff}}{N_{\rm in}} \frac{\sum_{i} \sigma_i^4}{\sum_{i} \sigma_i^2}
    \leq \frac{N_{\rm eff}}{N_{\rm in}}. 
    \label{eq:TE}
\end{equation}
Here, $N_{\rm eff} = (\sum_{i=1} \sigma_i^2)^2/\sum_{i=1} \sigma_i^4$ is the effective number of high-transmission channels characterized through an inverse participation ratio~\cite{2017_Hsu_nphys}, and it is independent of the overall prefactor of the transmission matrix.
Figure~\ref{fig:Fig1}(d) shows the normalized singular values of the transmission matrix $\mathbf{t}(k_y,k_y')$ in Fig.~\ref{fig:Fig1}(c), for which $\langle T \rangle \leq {N_{\rm eff}}/{N_{\rm in}} \approx 74$\%.

This ${N_{\rm eff}}$ comes directly from the desired angle-dependent response encoded in the transmission matrix, regardless of what structural design and material composition are used to realize such response.
Together with energy conservation, ${N_{\rm eff}}$ imposes a fundamental bound on the average transmission of the system through \eqref{eq:TE}.
This formalism applies to any linear optical system for which the desired response is known.

A bound is useful only when it is sufficiently tight.
Supplementary Fig.~S2 shows from full-wave numerical simulations~\cite{2022_Lin_NCS} that Eq.~(\ref{eq:TE}) indeed provides a reasonably tight upper bound for the averaged transmission efficiency $\langle T \rangle$, considering hyperbolic and quadratic metalenses as examples.

Figure~\ref{fig:Fig2}(a) shows the transmission efficiency bound $N_{\rm eff}/N_{\rm in}$ as a function of NA for aberration-free nonlocal metalenses with FOV larger than the threshold shown in Supplementary Fig.~S1.
Since FOV and the output diameter $D_{\rm out}$ have a very small influence on $N_{\rm eff}/N_{\rm in}$ (Supplementary Figs.~S3,S4), in Fig.~\ref{fig:Fig2}(a) we map out how the efficiency bound depends on the NA while averaging over FOV and $D_{\rm out}$.
We see that using equal entrance and output diameters, {\it i.e.}, $D_{\rm in}=D_{\rm out}$, results in an efficiency bound that drops approximately as $\sqrt{1-{\rm NA}^2}$. This bound applies regardless of how complicated or optimized the design is and regardless of what materials are used.

In Fig.~\ref{fig:Fig2}(b,c), we show the distribution (among different FOV and output diameters) of the transmission efficiency bound with NA = 0.7 and 0.9.
The bound is consistent with the inverse design results of Ref.~\cite{lin2021computational}, where the achieved average absolute focusing efficiency (considering only the transmitted power within three full-widths at half-maximum around the focal peak) is 25\% for a nonlocal metalens with NA = 0.7 and FOV = 80$^{\circ}$.

\subsection{Optimal aperture size}

\begin{figure}[tb]
\centering
\includegraphics[width=0.28\textwidth]{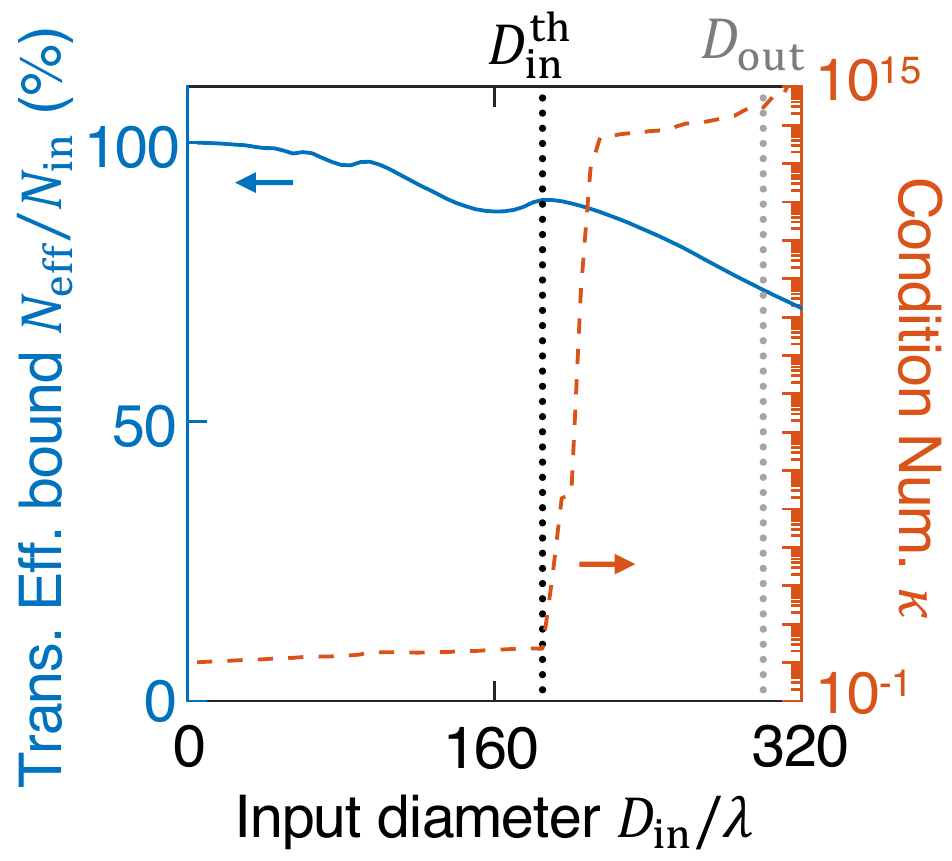}
\caption{
The transmission efficiency bound $N_{\rm eff}/N_{\rm in}$ and the condition number $\kappa$ of the ideal transmission matrix as a function of the entrance diameter $D_{\rm in}$.
Black and gray vertical dotted lines indicate $D_{\rm in}^{\rm th}=185\lambda$ and $D_{\rm out}=300\lambda$.
Lens parameters: output diameter $D_{\rm out}=300\lambda$, NA = 0.8, FOV = 140$^{\circ}$.}
\label{fig:Fig3}
\end{figure}

In addition to establishing a transmission efficiency limit, it would be even more useful to know what strategy one may adopt to raise such an efficiency limit.
To this end, we examine the singular values of the transmission matrix.
In Fig.~\ref{fig:Fig1}(d) where $D_{\rm in}=D_{\rm out}=300\lambda$ and NA = 0.8, 
we see there are many zero singular values. These zero singular values lower $N_{\rm eff}$ and the transmission efficiency. 
Removing these zero singular values ({\it i.e.}, eliminating non-transmitting wavefronts) can raise the transmission efficiency bound $N_{\rm eff}/N_{\rm in}$.
A zero singular value corresponds to a superposition of the columns of the transmission matrix that yields a zero vector, meaning those columns are linearly dependent.
Therefore, we can eliminate zero singular values by reducing the number of columns in the transmission matrix. Because the input wave vectors $|k_y'|<(2\pi/\lambda)\sin{(\rm FOV/2)}$ are sampled with momentum spacing $2\pi/D_{\rm in}$ at the Nyquist rate, we expect that reducing $D_{\rm in}$ can lower the number of input columns in the transmission matrix to raise the transmission efficiency bound.
Figure~\ref{fig:Fig3} shows this strategy indeed works: reducing the input aperture size $D_{\rm in}$ increases the efficiency bound $N_{\rm eff}/N_{\rm in}$. 

\begin{figure*}[ht]
\centering
\includegraphics[width=0.98\textwidth]{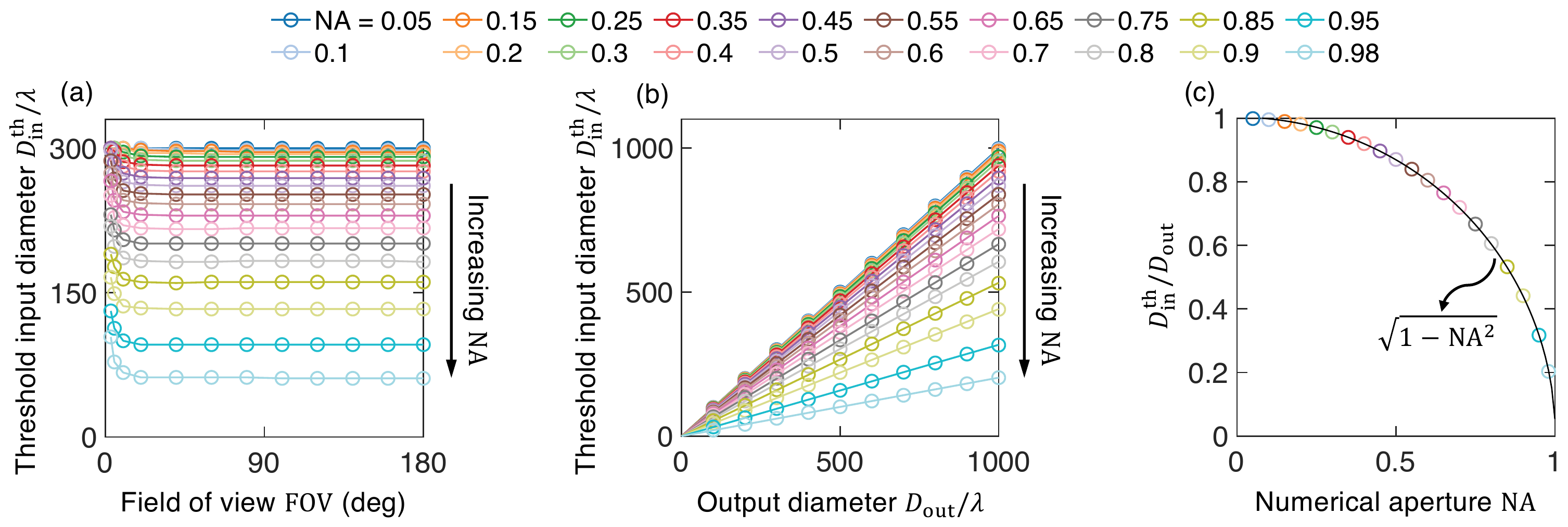}
\caption{Dependence of the threshold input diameter $D_{\rm in}^{\rm th}$ on the lens parameters.
(a) $D_{\rm in}^{\rm th}$ as a function of the FOV when $D_{\rm out}=300\lambda$.
(b) $D_{\rm in}^{\rm th}$ as a function of the output diameter $D_{\rm out}$ when FOV = 140$^{\circ}$. Symbols are $D_{\rm in}^{\rm th}$ computed from the ideal transmission matrices, and solid lines are linear fits.
(c) $D_{\rm in}^{\rm th}/D_{\rm out}$ averaged over different FOV for nonlocal metalenses as a function of NA. Black solid line is $\sqrt{\rm 1-NA^2}$.}
\label{fig:Fig4}
\end{figure*}

The next question is: what would be an optimal input diameter $D_{\rm in}$ to use? While reducing $D_{\rm in}$ raises the transmission efficiency bound, doing so also reduces the amount of light that can enter the metalens, which is not desirable.
To find a balance, we examine
the condition number $\kappa$, defined as the ratio between the maximal and the minimal singular values.
When zero $\sigma_i$ exist, the condition number $\kappa$ diverges to infinity (subject to numerical precision).
The right-axis of Fig.~\ref{fig:Fig3} shows that $\kappa$ abruptly shoots up by many orders of magnitude when the input diameter $D_{\rm in}$ raises above a threshold that we label as $D_{\rm in}^{\rm th}$ (black dotted line).
When $D_{\rm in} < D_{\rm in}^{\rm th}$, $\kappa$ is of order unity, all singular values are comparable with no zero-transmission wavefronts, so we have ${N_{\rm eff}} \approx N_{\rm in}$, and the transmission efficiency bound is close to unity.
When $D_{\rm in} > D_{\rm in}^{\rm th}$, near-zero-transmission wavefronts start to appear, which results in a fast reduction of the transmission efficiency bound.
Therefore, this threshold value $D_{\rm in}^{\rm th}$ is an optimal input diameter to use, providing maximal entrance flux while keeping a near-unity transmission efficiency bound.

To automate the determination of $D_{\rm in}^{\rm th}$, we examine the slope $\partial \kappa/\partial D_{\rm in}$, which transitions from near zero to a very large number at $D_{\rm in}^{\rm th}$.
Supplementary Tab.~S1 shows that different threshold values for $\partial \kappa/\partial D_{\rm in}$ yield almost identical $D_{\rm in}^{\rm th}$ and $N_{\rm eff}/N_{\rm in}$.
Note that the choice of the global phase $\psi(\theta_{\rm in})$ in \eqref{eq:phase_2D} does not influence $D_{\rm in}^{\rm th}$ and the transmission efficiency bound, as shown in Supplementary Fig.~S5.

To guide future designs, it is desirable to know how $D_{\rm in}^{\rm th}$ depends on the various lens parameters. Figure~\ref{fig:Fig4} plots $D_{\rm in}^{\rm th}$ as a function of the FOV, output diameter $D_{\rm out}$, and NA.
As described in Sec.~2\ref{subsec:nonlocality}, lenses with a very small FOV do not require nonlocality; for such local metalenses, we find $D_{\rm in}^{\rm th} \approx D_{\rm out}$ as expected from the schematic in Fig.~\ref{fig:Fig1}(a).
As the FOV increases and nonlocality emerges, we observe in Fig.~\ref{fig:Fig4}(a) that $D_{\rm in}^{\rm th}$ drops below $D_{\rm out}$ (as expected from the preceding discussions) and reaches a constant value that depends on NA but not on the FOV.  
This $D_{\rm in}^{\rm th}$ for nonlocal metalenses is proportional to $D_{\rm out}$ [Fig.~\ref{fig:Fig4}(b)].
Figure~\ref{fig:Fig4}(a,b) fix $D_{\rm out}=300\lambda$ and FOV = 140$^{\circ}$ respectively; other lens parameters share similar dependencies (Figs.~S6-S7 of Supplement 1).
In Fig.~\ref{fig:Fig4}(c), we find empirically that the NA dependence is well described by
\begin{equation}
    D_{\rm in}^{\rm th} = D_{\rm out}\sqrt{1-{\rm NA}^2}.
    \label{eq:Din_th}
\end{equation}
This result provides a recipe for choosing the input and output aperture sizes for nonlocal high-NA metalenses with high transmission.

To demonstrate the increased transmission efficiency bound, we also show in Fig.~\ref{fig:Fig2}(a--c) the transmission efficiency bound when the input aperture size $D_{\rm in}$ is set to the optimal $D_{\rm in}^{\rm th}$ in \eqref{eq:Din_th}.
We observe a large transmission efficiency bound ${N_{\rm eff}}/{N_{\rm in}}$ that overcomes the $\sqrt{1-{\rm NA}^2}$ limit when equal entrance and output apertures are used.

\section{Discussion}
\label{sec:conclusion}

While this work was originally motivated by the efficiency of high-NA nonlocal metalenses, the formalism we introduce is very general.
Given the desired response of any multi-channel optical system, one may write down its transmission matrix and apply \eqref{eq:TE} to establish a bound on its transmission efficiency.
Like the thickness bound introduced in Refs.~\cite{li2022thickness, 2023_Miller_Science}, the transmission efficiency bound here is functionality-driven and design-independent.



Nonlocal metasurfaces open up a wide range of applications and tailored angle-dependent responses that are impossible for traditional local metasurfaces~\cite{shastri2022nonlocal}.
The efficiency bound in this work provides valuable guidance for this rapidly evolving field.

\section*{Acknowledgments}
We thank X.~Gao for helpful discussions.
This work is supported by the National Science Foundation CAREER award (ECCS-2146021) and the Sony Research Award Program.
\bf{Disclosures:} \rm The authors declare no conflicts of interest.
\bf{Data availability:} \rm All data needed to evaluate the conclusions in this study are presented in the paper and in the supplemental document.

\bibliography{sample}

\end{document}


\maketitle

\tableofcontents

\section{Nonlocality of metalenses}
\label{sec:local_nonlocal}

When a metalens achieves diffraction-limited focusing across a wide range of incident angles, its response must be angle-dependent, requiring nonlocality.
Conversely, if the angular range is small, the response does not need to vary with angle, and a local metalens would suffice.
Thus we expect a threshold field of view (FOV) that separates local and nonlocal metalenses.

Assume free space on the incident and transmitted sides of the metalens.
An ideal lens needs to match the optical path lengths of the marginal rays and the chief ray from the focal spot position $(y=f\tan{\theta_{\rm in}}, z=h+f)$ to the lens surface $(y,z=h)$ for diffraction-limited focusing, so the phase distribution on the back surface of an ideal metalens is
\begin{equation}
    \phi_{\rm out}^{\rm ideal}(y,z=h,\theta_{\rm in}) =
     \psi(\theta_{\rm in})-\frac{2\pi}{\lambda}\sqrt{f^2+\left (y-f\tan{\theta_{\rm in}} \right )^2},
     \label{eq:phase_2D}
\end{equation}
where $h$, $\theta_{\rm in}$ and $f$ are the lens thickness, incident angle and focal length, respectively. 
The phase shift provided by the metalens is then
\begin{equation}
\Delta \phi_{\rm ideal}(y,\theta_{\rm in})=\phi_{\rm out}^{\rm ideal}(y,\theta_{\rm in})-\phi_{\rm in}(y,\theta_{\rm in}),
\end{equation}
where $\phi_{\rm in}(y,\theta_{\rm in}) = k_y^{\rm in} y = (2\pi/\lambda)\sin{\theta_{\rm in}}y$ is the phase profile of the incident light.

The $\psi(\theta_{\rm in})$ in \eqref{eq:phase_2D} is an angle-dependent but spatially-invariant global phase, with no influence on the focusing performance.
One sensible choice for the global phase $\psi(\theta_{\rm in})$, applied in all calculations in this work, is
\begin{equation}
    \psi(\theta_{\rm in})= \frac{2\pi}{\lambda} \left \langle \sqrt{f^2+(y-f\tan{\theta_{\rm in}})^2}+y\sin{\theta_{\rm in}} \right \rangle_y \equiv \psi_0(\theta_{\rm in}),
\label{eq:psi_0}
\end{equation}
where $\langle \cdots \rangle_y$ denotes averaging over $y$ within the output aperture {\it i.e.}, $|y|<D_{\rm out}/2$. With this $\psi_0(\theta_{\rm in})$,
the $y$ average of the phase shift, $\langle \Delta \phi_{\rm ideal}(y,\theta_{\rm in})\rangle_y$, is the same for different incident angles, which minimizes the required thickness of the metalens~\cite{li2022thickness}.
We will show in Sec.~\ref{sec:psi} that the transmission efficiency bound studied in this work is independent of the choice of $\psi(\theta_{\rm in})$.

A local hyperbolic metalens achieves diffraction-limited focusing at the normal incidence $\theta_{\rm in}=0^{\circ}$, with an angle-independent phase-shift profile~\cite{hecht2017optics,aieta2012aberration}
\begin{equation}
    \Delta \phi_{\rm hyp}(y) = \Delta \phi_{\rm ideal}(y,\theta_{\rm in}=0^{\circ}) = \psi_0(\theta_{\rm in}=0^{\circ})-\frac{2\pi}{\lambda}\sqrt{f^2+y^2}.
    \label{eq:hyperbolic}
\end{equation}
Away from the normal incident angle, $\Delta \phi_{\rm hyp}(y) \neq \Delta \phi_{\rm ideal}(y,\theta_{\rm in})$, so a hyperbolic metalens no longer achieves ideal focusing.

Given the lens parameters (FOV, NA, and $D_{\rm out}$), we want to determine whether it can be realized with a local metalens or whether nonlocality is required.
To do so, we evaluate
\begin{equation}
    {\rm max}_{y,\theta_{\rm in}}\ |\Delta\phi_{\rm ideal}(y,\theta_{\rm in}) - \Delta\phi_{\rm hyp}(y)|
    \label{eq:FOV_threshold}
\end{equation}
across all positions $|y|<D_{\rm out}/2$ and all incident angles within $|\theta_{\rm in}|<{\rm FOV}/2$.
If Eq.~(\ref{eq:FOV_threshold})$<\pi$, we consider a local hyperbolic metalens to be sufficient, and the metalens is classified as local. 
Otherwise, it is classified as nonlocal. 

\begin{figure}[htbp]
\centering
\includegraphics[width=0.6\textwidth]{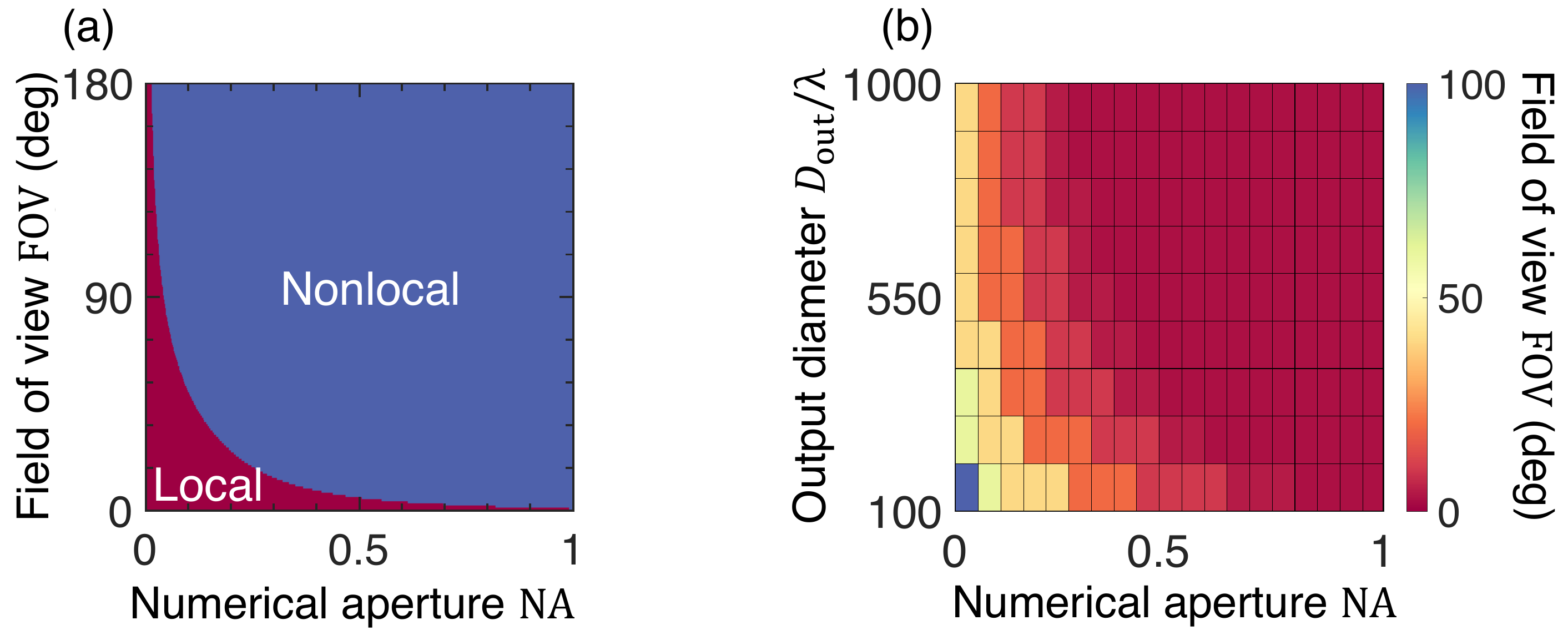}
\caption{(a) Designation of whether a metalens is local or nonlocal; $D_{\rm out}=300\lambda$. (b) The threshold FOV that separates local and nonlocal metalenses for different output diameter $D_{\rm out}$ and NA.}
\label{fig:nonlocal_local}
\end{figure}

Figure~\ref{fig:nonlocal_local}(a) shows whether an ideal metalens is local or nonlocal over $\rm FOV \in (0^{\circ},180^{\circ})$ and numerical aperture $\rm NA \in (0,1)$ when $D_{\rm out}=300\lambda$.
When the FOV is very small, all lenses are local regardless of NA; this is the reason that in Fig.~4(a) of the main text, $D_{\rm in}^{\rm th} \approx D_{\rm out}$ when the FOV is very small.
Above a threshold FOV, the lens becomes nonlocal, for which $D_{\rm in}^{\rm th} < D_{\rm out}$.
The threshold FOV and $D_{\rm in}^{\rm th}$ both depend on the NA.
Figure~\ref{fig:nonlocal_local}(b) further shows the threshold FOV for different $D_{\rm out}$ and NA.

\section{Transmission matrix of an ideal metalens}
\label{sec:transmission_matrix}

Following Eq.~(2) of the main text, the transmission matrix relates the incoming wavefront to the outgoing wavefront as $\beta_b = \sum_{a}t_{ba}\alpha_{a}$.

We express the input wavefront in a flux-orthogonal basis of truncated plane waves $\{f_a\}$,
\begin{equation}
   f_a(y,z)=\begin{cases}
   \frac{1}{\sqrt{D_{\rm in}}}\frac{1}{\sqrt{k_z^a}}e^{i[k_y^ay+k_z^az)]}  & \text{for } |y|<\frac{D_{\rm in}}{2}\\
   0  & \text{otherwise}
   \end{cases}
\label{eq:in_basis}
\end{equation}
with
$\{k_y^a\} = \left\{ a(2\pi/D_{\rm in}) \text{ such that } a\in \mathbb{Z} \text{ and } |k_y^a|<(2\pi/\lambda) \sin{(\rm FOV/2)}\right\}$ and $(k_y^a)^2+(k_z^a)^2=(2\pi/\lambda)^2$.

Consider incident plane wave from a fixed angle $\theta_{\rm in}^a$ within the FOV.
Here, $E_x^a(y',z=0)=f_a(y',z=0)$, corresponding to one column of the transmission matrix, so $\beta_b = t_{ba}$.
To perfectly focus to the focal spot, the field profile on the output surface ($z=h$) of the metalens should be proportional to the conjugated field radiated from a point source at the focal spot.
Therefore,
\begin{equation}
E_x^a(y,z=h)=\begin{cases}
A(\theta_{\rm in}^a)\frac{e^{i\phi_{\rm out}^{\rm ideal}(y,\theta_{\rm in}^a)}}{[f^2+(y-f\tan{\theta_{\rm in}^a})^2]^{{1}/{4}}} & \text{for } |y|<\frac{D_{\rm out}}{2} \\
0 & \text{otherwise}
\end{cases}.
\label{eq:Ez_at_h}
\end{equation}
The distance between a point $(y,z=h)$ on the back surface of the metalens and the focal spot $(y=f\tan{\theta_{\rm in}},z=h+f)$ is $r=\sqrt{f^2+(y-f\tan{\theta_{\rm in}})^2}$.
The amplitude factor $1/\sqrt{r}$ comes from the decay rate of the radiated field from a point source in 2D.
We expand this ideal output in a basis of flux-orthogonal truncated plane waves, 
\begin{equation}
   g_b(y,z)=\begin{cases}
   \frac{1}{\sqrt{D_{\rm out}}}\frac{1}{\sqrt{k_z^b}}e^{i[k_y^by+k_z^b(z-h)]}  & \text{for } |y|<\frac{D_{\rm out}}{2}\\
   0  & \text{otherwise}
   \end{cases}
\label{eq:out_basis}
\end{equation}
with
$\{k_y^b\} = \left\{ b(2\pi/D_{\rm out}) \text{ such that } b\in \mathbb{Z} \text{ and } |k_y^b|<2\pi/\lambda \right\}$ and $(k_y^b)^2+(k_z^b)^2=(2\pi/\lambda)^2$.
Projecting onto this basis, we obtain $E_x(y,z=h)=\sum_b\beta_bg_b(y,z=h)$ with
\begin{equation}
    t_{ba} = \beta_b = \sqrt{\frac{k_z^b}{D_{\rm out}}} \int_{-\frac{D_{\rm out}}{2}}^{\frac{D_{\rm out}}{2}}E_x^a(y,z=h)e^{-ik_y^by}dy.
\label{eq:t_ba}
\end{equation}

In practice, we can approximate the continuous integration over $y$ in Eq.~(\ref{eq:t_ba}) by a discrete summation. A sampling spacing of $\Delta y = \lambda/2$ is chosen following the Nyquist-Shannon sampling theorem~\cite{landau1967sampling}.
The discretized Eq.~(\ref{eq:t_ba}) can be evaluated efficiently using fast Fourier transform:
\begin{equation}
    t_{ba} \approx \Delta y \sqrt{\frac{k_z^b}{D_{\rm out}}} 
    e^{-ik_y^b(-\frac{D_{\rm out}}{2}+\frac{\Delta y}{2})}\sum_{n=0}^{N-1}E_x^a(y_n,z=h)e^{-i\frac{2\pi}{N}b n},
\label{eq:t_ba_fft}
\end{equation}
where $y\in [-D_{\rm out}/2,D_{\rm out}/2]$ is discretized to $\{y_n\equiv -\frac{D_{\rm out}}{2}+\frac{\Delta y}{2}+n\Delta y\}$, and $N\equiv D_{\rm out}/\Delta y\in \mathbb{Z}$ is the length of $\{y_n\}$.

Eq.~(\ref{eq:Ez_at_h}) includes an amplitude factor $A(\theta_{\rm in}^a)$ that can depend on the incident angle.
This means that each column of the transmission matrix has an undetermined amplitude prefactor.
While the average transmission efficiency bound $N_{\rm eff}/N_{\rm in}$ in this work does not depend on a global amplitude prefactor, we do need to specify the relative amplitude between the columns.
Here, we choose $A(\theta_{\rm in}^a)$ such that $\sum_b |t_{ba}|^2$ is the same for all incident angles within the FOV.
The reason is twofold. First, it is typically desirable that the focal power is independent of the incident angle so there is no vignetting in the image, and the focal power is proportional to the transmitted flux $\sum_b |t_{ba}|^2$ times the Strehl ratio (which is unity for an ideal lens).
Second, in this work we are interested in the upper bound on the transmission efficiency, so we want to find the largest possible $N_{\rm eff}$. A large $N_{\rm eff}$ corresponds to all possible inputs having similar total transmission, so making $\sum_b |t_{ba}|^2$ independent of the incident angle can increase $N_{\rm eff}$.

\section{Transmission efficiency of hyperbolic and quadratic metalenses}
\label{sec:TE_definition}

\begin{figure}[htbp]
\centering
\includegraphics[width=0.55\textwidth]{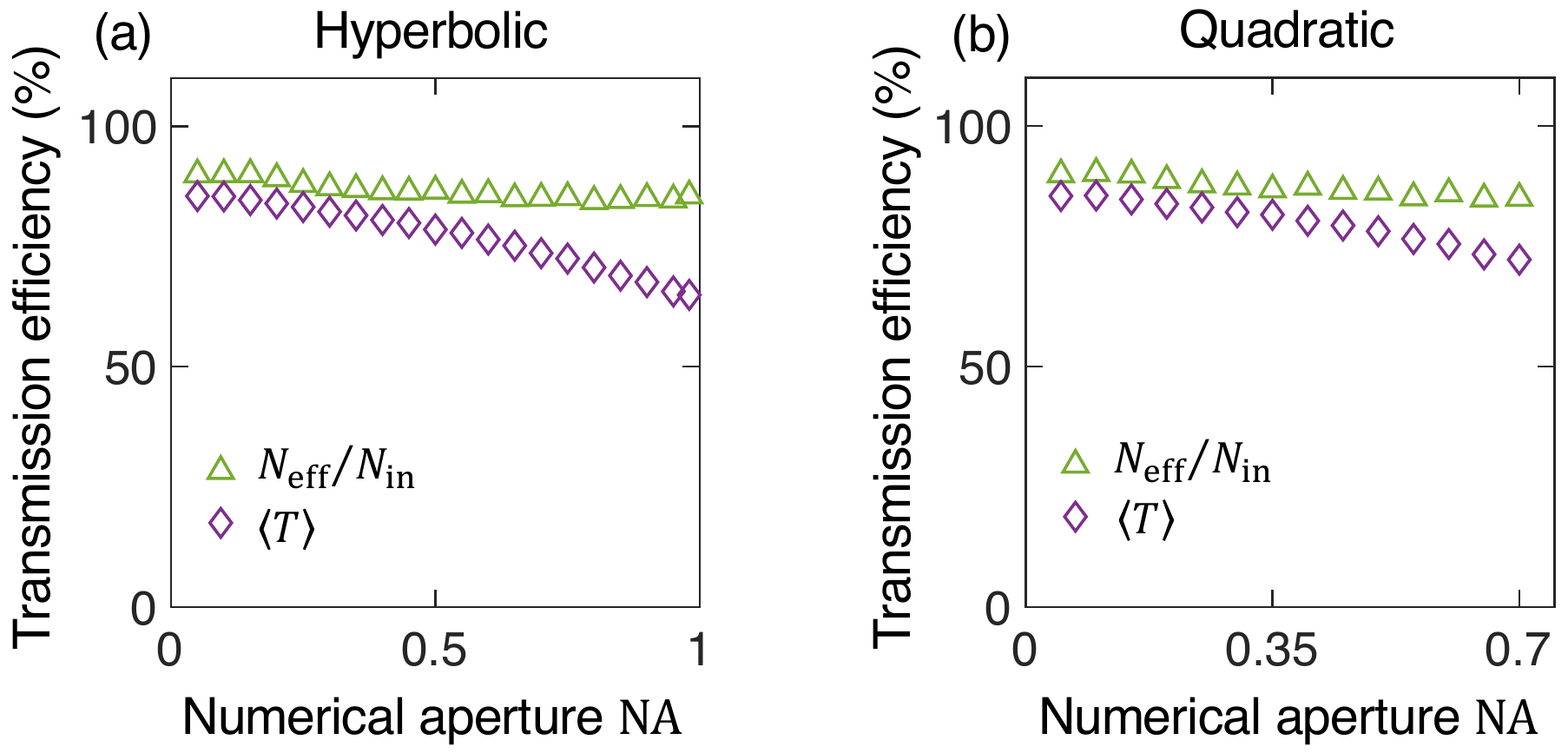}
\caption{Transmission efficiency for 2D metalenses with (a) hyperbolic phase-shift profile and (b) quadratic phase-shift profile. Lens parameters: diameter $D$ = 100 \textmu m, $\lambda=532$ nm.}
\label{fig:TE_of_real}
\end{figure}

As a verification, here we compare the transmission efficiency bound $N_{\rm eff}/N_{\rm in}$ with the transmission efficiency of actual metalenses.

We design 2D metalenses with hyperbolic [Eq.~(\ref{eq:hyperbolic})]
and quadratic~\cite{pu2017nanoapertures,martins2020metalenses,2021_Lassalle_ACSph} phase-shift profiles
\begin{equation}
    \Delta \phi_{\rm qua}(y) = -\frac{2\pi}{\lambda}\frac{y^2}{2f},
    \label{eq:quadratic}
\end{equation}
operating at wavelength $\lambda=532$ nm, composed of ridges with thickness $h=0.6$ \textmu m and varying widths between 45 nm and 200 nm.
Each unit cell has a titanium dioxide ridge with refractive index $n_{\rm ridge}=2.43$ sitting on a silica substrate with $n_{\rm sub}=1.46$.
The unit cell size is fixed at 240 nm.
We calculate their transmission matrices by full-wave simulations using an open-source software MESTI~\cite{MESTI,2022_Lin}.
The simulation domain is discretized with 40 pixels per wavelength. 

The transmission efficiency $T_a$ at incident angle $\theta_{\rm in}^a$ is
\begin{equation}
    T_a = \sum_b |t_{ba}|^2
    \label{eq:TE_a}
\end{equation}
for a flux-normalized transmission matrix $t_{ba}$.
Here we consider transmission averaged over incident angles with $|\theta_{\rm in}^a|<45^{\circ}$ ({\it ie.} $\langle T \rangle=\sum_{b,a}|t_{ba}|^2/N_{\rm in}$) and compare it to the efficiency bound $N_{\rm eff}/N_{\rm in}$.

Figure~\ref{fig:TE_of_real} plots $N_{\rm eff}/N_{\rm in}$ and $\langle T \rangle$ as a function of NA with the lens diameter fixed at $D=100$ \textmu m, for hyperbolic and quadratic metalenses.
Indeed, $\langle T \rangle$ is always below $N_{\rm eff}/N_{\rm in}$. 
Note that the range of NA is restricted to $<0.71$ for the quadratic metalens because its largest possible effective NA is $1/\sqrt{2} = 0.71$~\cite{2021_Lassalle_ACSph}.

\section{Dependence of transmission efficiency bound on lens parameters}
\label{sec:TE}

\begin{figure}[htbp]
\centering
\includegraphics[width=0.6\textwidth]{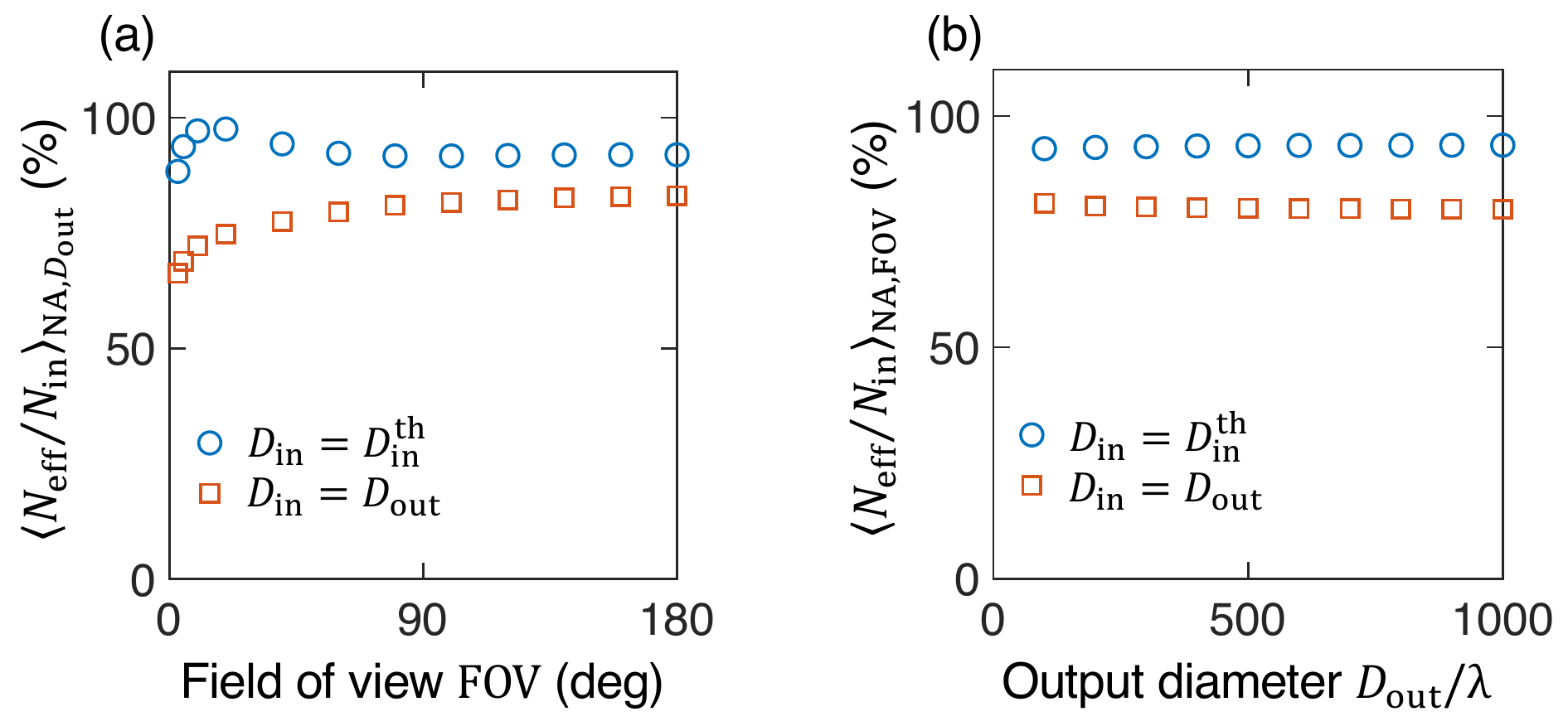}
\caption{Transmission efficiency bound $N_{\rm eff}/N_{\rm in}$ of nonlocal metalenses (a) averaged over NA and output diameter $D_{\rm out}$ as a function of the FOV,
and (b) averaged over NA and FOV as a function of the output diameter $D_{\rm out}$, when $D_{\rm in}=D_{\rm in}^{\rm th}$ and $D_{\rm in}=D_{\rm out}$.}
\label{fig:TE_averaged}
\end{figure}

Figure~\ref{fig:TE_averaged} plots the transmission efficiency bound $N_{\rm eff}/N_{\rm in}$ as a function of the FOV and $D_{\rm out}$ when $D_{\rm in}=D_{\rm in}^{\rm th}$ and $D_{\rm in}=D_{\rm out}$, respectively, averaging over the other lens parameters.
We see that FOV and $D_{\rm out}$ have very minor effects on the efficiency bound.

\begin{figure}[htbp]
\centering
\includegraphics[width=0.9\textwidth]{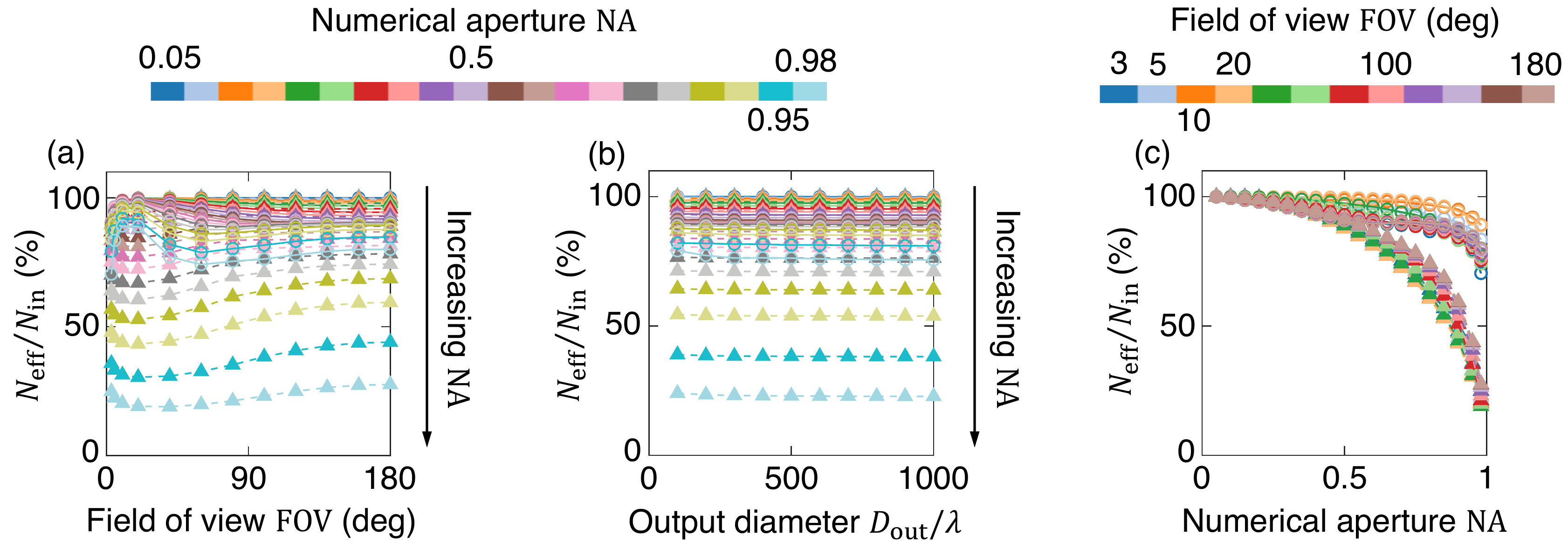}
\caption{Transmission efficiency bound $N_{\rm eff}/N_{\rm in}$ of nonlocal metalenses (a) as a function of the FOV when $D_{\rm out}=300\lambda$, (b) as a function of the output diameter $D_{\rm out}$ when FOV = 80$^{\circ}$, and (c) as a function of the NA when $D_{\rm out}=300\lambda$. Open circles and filled triangles represent the bound with $D_{\rm in}=D_{\rm in}^{\rm th}$ and $D_{\rm in}=D_{\rm out}$, respectively.}
\label{fig:TE}
\end{figure}

Some representative results of the transmission efficiency bound are plotted in Fig.~\ref{fig:TE}.

\section{Threshold input diameter from $\partial \kappa/\partial D_{\rm in}$}
\label{sec:threshold}

Table~\ref{tab:tab1} shows the threshold input diameter $D_{\rm in}^{\rm th}$ for high transmission determined by $\left(\partial \kappa/\partial D_{\rm in}\right)\vert_{D_{\rm in} = D_{\rm in}^{\rm th}}=0.04/\lambda$, 0.4$/\lambda$ or 4$/\lambda$, and the corresponding transmission efficiency bound $N_{\rm eff}/N_{\rm in}$ at $D_{\rm in}=D_{\rm in}^{\rm th}$ when $D_{\rm out}=300\lambda$.
We can see that the results are insensitive to the threshold value used, indicating that $\partial \kappa/\partial D_{\rm in}$ is a robust indicator for the threshold.
We use $\left(\partial \kappa/\partial D_{\rm in}\right)\vert_{D_{\rm in} = D_{\rm in}^{\rm th}}=0.4/\lambda$ in the main text, though other values will yield similar results. 

\begin{table}[htbp]
\centering
\caption{\bf Threshold input diameter $D_{\rm in}^{\rm th}$ and transmission efficiency bound $N_{\rm eff}/N_{\rm in}$ at $D_{\rm out}=300\lambda$}
\begin{tabular}{lrrrrrr}
\toprule
   & \multicolumn{2}{c}{$\partial \kappa/\partial D_{\rm in}=0.04/\lambda$} & \multicolumn{2}{c}{$\partial \kappa/\partial D_{\rm in}=0.4/\lambda$} & \multicolumn{2}{c}{$\partial \kappa/\partial D_{\rm in}=4/\lambda$} \\
\cmidrule(ll){2-3} \cmidrule(ll){4-5} \cmidrule(ll){6-7}
   & \,\,\,\,\,\,\,$D_{\rm in}^{\rm th}$ &  $N_{\rm eff}/N_{\rm in}$\,\,\,\,\,\,\, &  \,\,\,\,\,\,\,$D_{\rm in}^{\rm th}$ & $N_{\rm eff}/N_{\rm in}$\,\,\,\,\,\,\, & \,\,\,\,\,\,\,$D_{\rm in}^{\rm th}$ & $N_{\rm eff}/N_{\rm in}$\,\,\,\,\,\,\, \\
\midrule
  NA = 0.2, FOV = 10$^{\circ}$ & 294$\lambda$ & 100\%\,\,\,\,\,\,\,\,\,\, & 300$\lambda$ & 98.7\%\,\,\,\,\,\,\,\,\, & 300$\lambda$ & 98.7\%\,\,\,\,\,\,\,\,\, \\
  NA = 0.5, FOV = 10$^{\circ}$ & 260$\lambda$ & 99.8\%\,\,\,\,\,\,\,\,\, & 265$\lambda$ & 98.8\%\,\,\,\,\,\,\,\,\, & 274$\lambda$ & 95.9\%\,\,\,\,\,\,\,\,\, \\
  NA = 0.9, FOV = 10$^{\circ}$ & 133$\lambda$ & 96.0\%\,\,\,\,\,\,\,\,\, & 136$\lambda$ & 95.0\%\,\,\,\,\,\,\,\,\, & 146$\lambda$ & 89.3\%\,\,\,\,\,\,\,\,\, \\
  NA = 0.2, FOV = 80$^{\circ}$ & 293$\lambda$ & 98.4\%\,\,\,\,\,\,\,\,\, & 294$\lambda$ & 98.7\%\,\,\,\,\,\,\,\,\, & 296$\lambda$ & 99.0\%\,\,\,\,\,\,\,\,\, \\
  NA = 0.5, FOV = 80$^{\circ}$ & 260$\lambda$ & 92.8\%\,\,\,\,\,\,\,\,\, & 261$\lambda$ & 93.0\%\,\,\,\,\,\,\,\,\, & 264$\lambda$ & 93.4\%\,\,\,\,\,\,\,\,\, \\
  NA = 0.9, FOV = 80$^{\circ}$ & 133$\lambda$ & 84.2\%\,\,\,\,\,\,\,\,\, & 133$\lambda$ & 84.2\%\,\,\,\,\,\,\,\,\, & 136$\lambda$ & 84.6\%\,\,\,\,\,\,\,\,\, \\
  NA = 0.2, FOV = 160$^{\circ}$ & 292$\lambda$ & 97.3\%\,\,\,\,\,\,\,\,\, & 294$\lambda$ & 97.9\%\,\,\,\,\,\,\,\,\, & 296$\lambda$ & 98.3\%\,\,\,\,\,\,\,\,\, \\
  NA = 0.5, FOV = 160$^{\circ}$ & 260$\lambda$ & 90.6\%\,\,\,\,\,\,\,\,\, & 261$\lambda$ & 90.8\%\,\,\,\,\,\,\,\,\, & 264$\lambda$ & 91.1\%\,\,\,\,\,\,\,\,\, \\
  NA = 0.9, FOV = 160$^{\circ}$ & 132$\lambda$ & 87.4\%\,\,\,\,\,\,\,\,\, & 133$\lambda$ & 87.7\%\,\,\,\,\,\,\,\,\, & 136$\lambda$ & 88.0\%\,\,\,\,\,\,\,\,\, \\
\bottomrule
\end{tabular}
  \label{tab:tab1}
\end{table}

\section{$\psi(\theta_{\rm in})$ dependence}
\label{sec:psi}

\begin{figure}[htbp]
\centering
\includegraphics[width=0.6\textwidth]{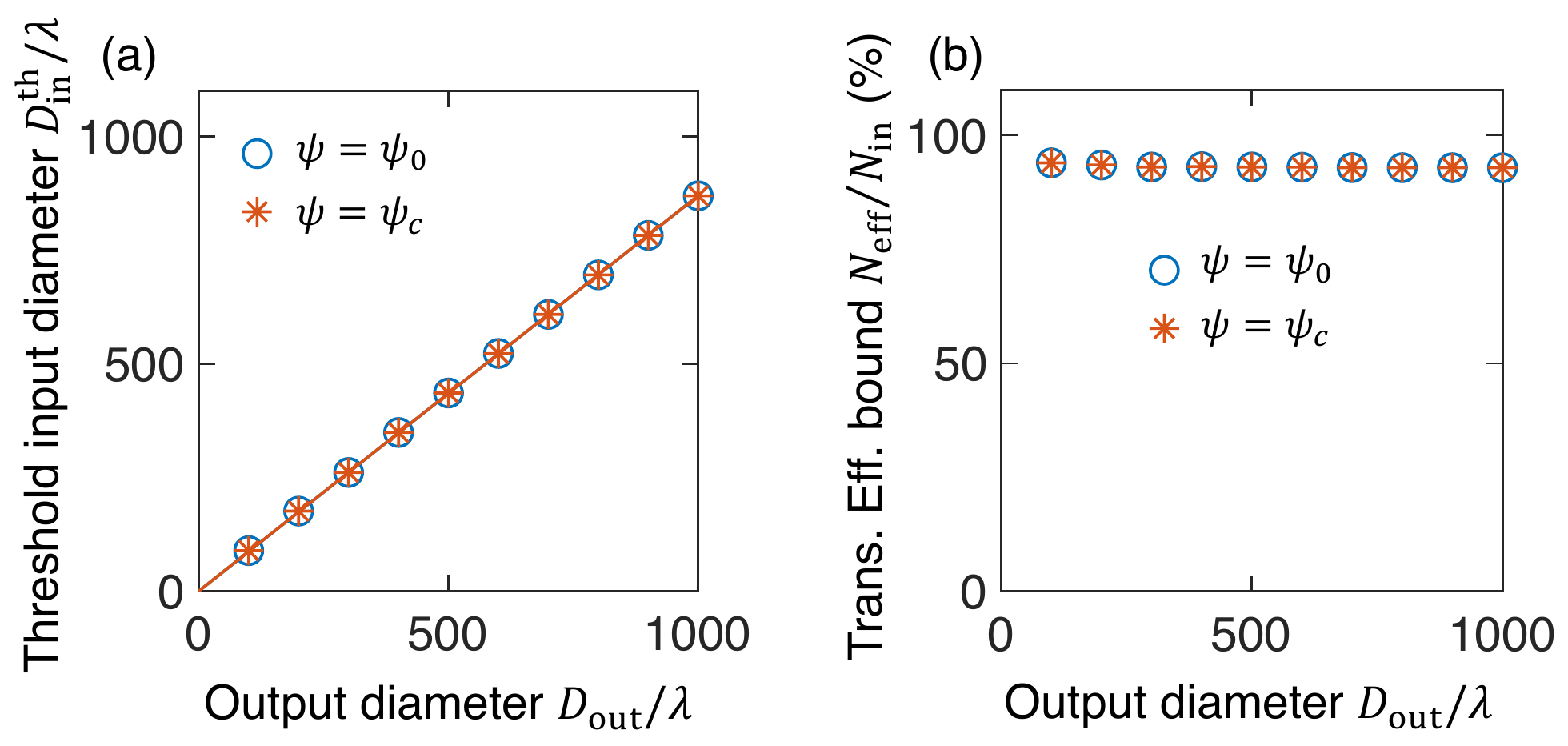}
\caption{Threshold input diameter $D_{\rm in}^{\rm th}$ and transmission efficiency bound $N_{\rm eff}/N_{\rm in}$ at $D_{\rm in}=D_{\rm in}^{\rm th}$ as a function of the output diameter $D_{\rm out}$ when different $\psi(\theta_{\rm in})$ is applied. Lens parameters: NA = 0.7, FOV = 80$^{\circ}$.}
\label{fig:generality}
\end{figure}

The incident-angle-dependent but spatially-invariant global phase $\psi(\theta_{\rm in})$ in Eq.~(\ref{eq:phase_2D}) has no influence on the focusing quality, so it can be chosen at will. 
The $\psi(\theta_{\rm in}) = \psi_0(\theta_{\rm in})$ in Eq.~(\ref{eq:psi_0}) was chosen to minimize the required thickness of the metalens~\cite{li2022thickness}. Another common choice is
\begin{equation}
    \psi(\theta_{\rm in})= \frac{2\pi}{\lambda} \sqrt{f^2+(f\tan{\theta_{\rm in}})^2} \equiv \psi_c(\theta_{\rm in}),
\label{eq:psi_c}
\end{equation}
which makes $\Delta \phi_{\rm ideal}(y=0,\theta_{\rm in})=\phi_{\rm out}^{\rm ideal}(y=0,\theta_{\rm in})-\phi_{\rm in}(y=0,\theta_{\rm in})=0$.

As shown in Fig.~\ref{fig:generality}, the threshold input diameter $D_{\rm in}^{\rm th}$ and transmission efficiency bound $N_{\rm eff}/N_{\rm in}$ at $D_{\rm in}=D_{\rm in}^{\rm th}$ are the same for both choices of the global phase $\psi(\theta_{\rm in})$.


\section{Comprehensive data on the threshold input diameter}
\label{sec:Din_th}

\begin{figure}[htbp]
\centering
\includegraphics[width=0.92\textwidth]{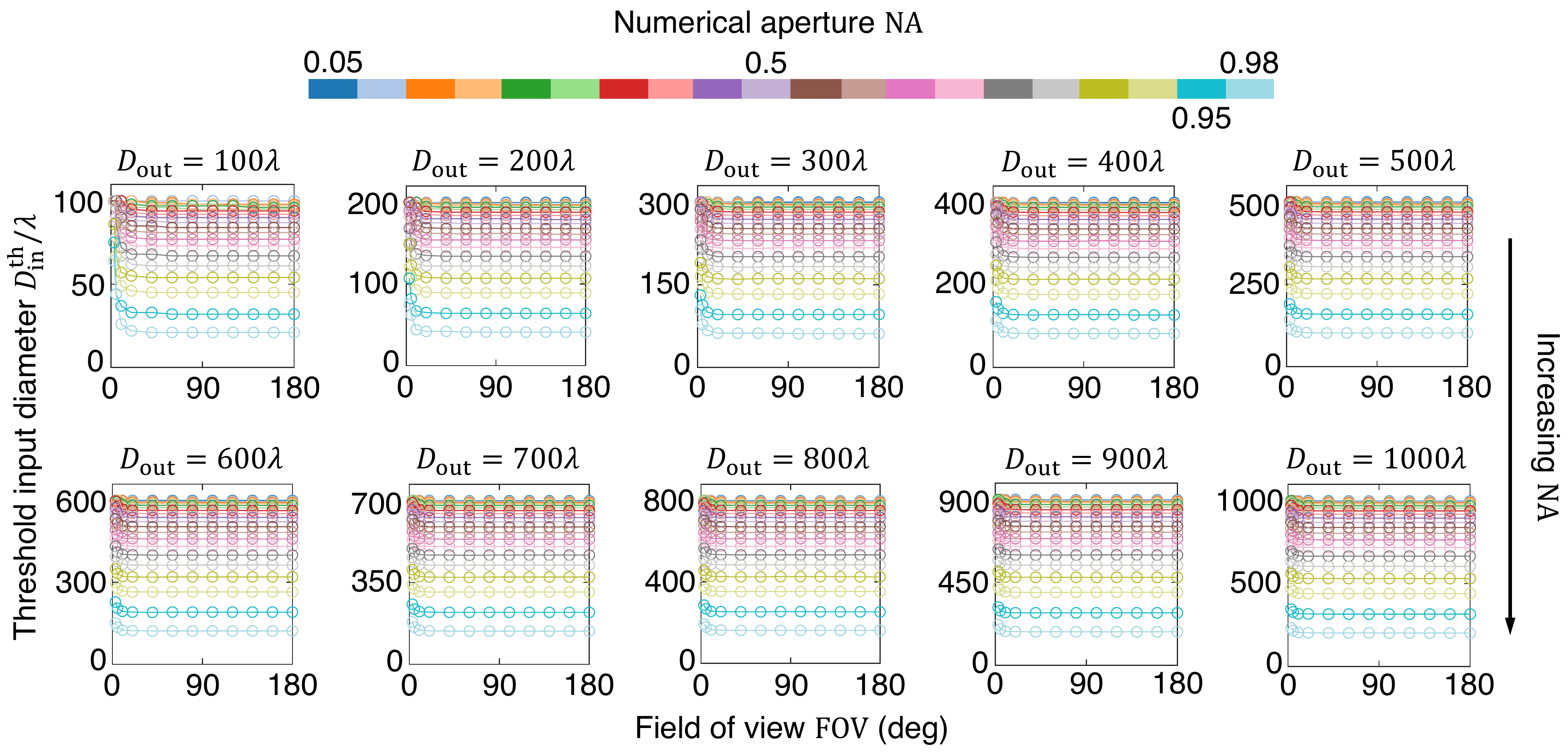}
\caption{Threshold input diameter $D_{\rm in}^{\rm th}$ as a function of the FOV for different lens diameter $D_{\rm out}$ and NA.}
\label{fig:Dinth_FOV}
\end{figure}

\begin{figure}[htbp]
\centering
\includegraphics[width=0.92\textwidth]{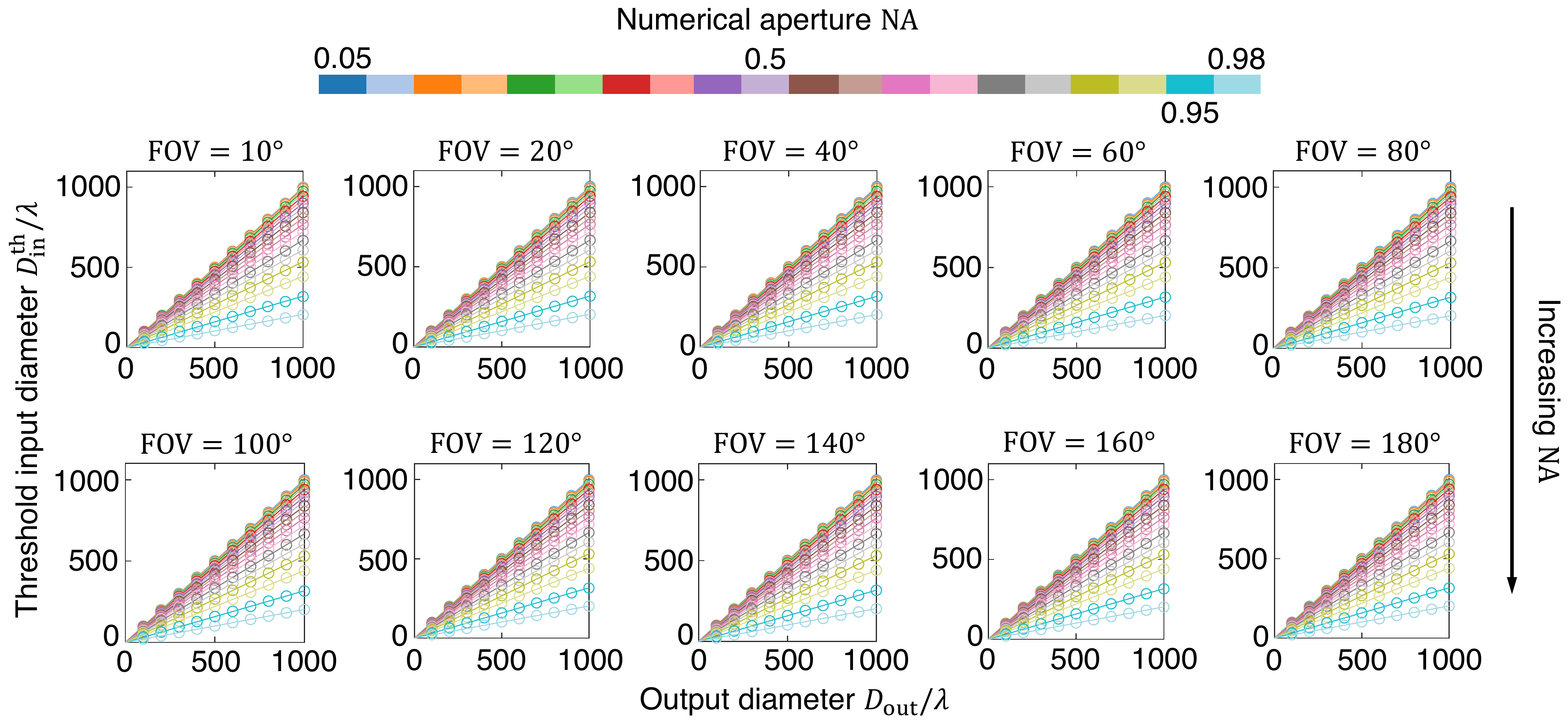}
\caption{Threshold input diameter $D_{\rm in}^{\rm th}$ as a function of the output diameter $D_{\rm out}$ for different FOV and NA.}
\label{fig:Dinth_Dout}
\end{figure}

Figure~4(a,b) of the main text plots the threshold input diameter $D_{\rm in}^{\rm th}$ for output diameter $D_{\rm out}=300\lambda$ and FOV = 140$^{\circ}$ respectively, with varying NA.
Figures~\ref{fig:Dinth_FOV}--\ref{fig:Dinth_Dout} plot $D_{\rm in}^{\rm th}$ for other $D_{\rm out}$ and other FOV.

\bibliography{supplemental-sample}
